# Dirac Equation and the Definition of Elementary Particles

## Jian-Hua Shu

*College of Information Science & Engineering, Huaqiao University, Quanzhou, Fujian 362021, China*

A concept of the total velocity that contains velocity and oscillatory velocity is proposed for the velocity solution of Dirac equation. It is shown that the electronic rest energy all comes from the oscillation of the electron itself. For this reason, the velocity solution of Dirac equation is taken as the definition of elementary particles. Leptons, mesons and baryons appear in results as the newly defined elementary particles, but the particle that consists of more than three quarks is ruled out. The results also show that a quark is not a particle, but part of the hadron or a partial particle and that quark confinement may serve as an evidence of this conclusion.

The negative energy solution is generally considered as the difficulty of Dirac equation. Nevertheless, it was undeniable that it was in the process of interpreting the negative energy solution that Dirac predicted the existence of positrons. We want to state that the negative energy solution is not entirely the fault of Dirac equation, because the information about particles is too much and complex for Dirac equation, which is a single-electron wave equation, to express it in a single way. In other words, the negative energy solution may be regarded as an indirect way in which the existence of positrons is expressed. Similarly, there is still the velocity solution of Dirac equation that is as inexplicable as the negative energy solution, because we can conclude that a measurement of a component of the velocity of a free electron is certain to lead to the result $\pm c$ from the velocity solution. We have never doubted the correctness of Dirac equation and believe that some unknown information about particles must have concealed behind the velocity solution. For this reason, we review the solution procedure of electronic velocity at first[1].

The free electron Dirac equation is

$$i\hbar \frac{\partial}{\partial t}\psi = H\psi$$

$$H = c\vec{\alpha} \cdot \vec{p} + \beta m_0 c^2 \qquad (1)$$

where

$$\alpha_j = \begin{bmatrix} \sigma_j & 0 \\ 0 & \sigma_j \end{bmatrix}, \quad \beta = \begin{bmatrix} I & 0 \\ 0 & -I \end{bmatrix} \qquad j = x, y, z \qquad (2)$$

$\sigma_j$ is the Pauli matrix and $I$ is a $2\times 2$ unit matrix. The $x$-component of the velocity is

$$\dot{x} = \frac{1}{i\hbar}[x, H] = c\alpha_x \qquad (3)$$

---

E-mail address: qrshu598@163.com

As $\dot{y}$ and $\dot{z}$ are similar, we can conclude that a measurement of a component of the velocity of a free electron is certain to lead to the result $\pm c$. This conclusion is easily seen to hold also when there is a field present. Let us now examine how the velocity of the electron varies with time. We have

$$i\hbar \dot{\alpha}_x = \alpha_x H - H\alpha_x \tag{4}$$

Since 
$$\alpha_x H + H\alpha_x = 2cp_x \tag{5}$$

and hence 
$$i\hbar \dot{\alpha}_x = 2\alpha_x H - 2cp_x$$

$$= 2cp_x - 2H\alpha_x \tag{6}$$

Finally, the $x$-component of an electronic velocity is expressed as

$$\dot{x} = c\alpha_x = v_{ox} + v_x \tag{7}$$

where

$$v_{ox} = \frac{i}{2}\hbar c \dot{\alpha}_x^0 e^{-i2Ht/\hbar} H^{-1} \tag{7a}$$

$$v_x = c^2 p_x H^{-1} \tag{7b}$$

The above is the partial process of solving electronic velocity.

In quantum mechanics, coordinates and momentum are generally taken as a pair of basic dynamic variables, others, such as angular momentum or kinetic energy, are expressed by them. In this sense, what represents the $x$-component of electronic velocity in Eq.(7) ought to be $v_x$ rather than $\dot{x}$, which is the classical definition of particle velocity and usually considered to be equivalent to $v_x$. Nevertheless, the fact is that $\dot{x}$ contains $v_{ox}$ called the $x$-component of oscillatory velocity besides $v_x$. We know that the definition $v_x = \dot{x}$ was given in classical mechanics, where a body is the main study object. Whether the definition is completely suitable to microcosmic particles is not known. According to the attitude of seeking truth from facts, we call $\dot{x}$ the $x$-component of the total velocity that contains velocity and oscillatory velocity for an electron for the time being. But as far as a body, which can be regarded as a composite particle because it also consists of many particles, is concerned, $\dot{x}$ still represents the $x$-component of particle velocity. In other words, a composite particle has no the concept of total velocity.

Square both members of Eq.(7) and make use of Eq.(6), we obtain

$$c^2 = v_{ox}^2 + v_x^2 \tag{8}$$

where

$$v_{ox}^2 = (\hbar c \dot{\alpha}_x^0 / 2)^2 H^{-2} \tag{8a}$$

$$v_x^2 = c^4 p_x^2 H^{-2} \tag{8b}$$

We notice that both members of Eq.(8) are asymmetrical, namely, the right one is associated with the coordinate direction, but the left is not. Besides, we notice that the electronic oscillatory velocity has some peculiar properties. It is summarized as follows:

① The oscillatory velocity contributes nothing to electronic momentum.
② The maximum of oscillatory velocity can achieve the velocity of light.

There are simply two explanations to the property ①. One is that the average oscillatory velocity through a time-interval equals to zero, but it is vetoed soon because the velocity, momentum and oscillatory velocity are all instantaneous values. The other is that oscillation is assumed to be isotropic. In reality, not only can the latter explain the property ①, but also ② because an isotropic oscillation would not bring a measurable move for the electron.

It is necessary to reconsider the physical meaning of $\dot{x}$ after the oscillation is assumed to be isotropic. The isotropy of oscillation implies that $v_{ox}$ has nothing to do with the direction and we call it the oscillatory speed. The left side of Eq.(8), as mentioned earlier, has nothing to do with the direction. However, $v_x$ is obviously concerned with the direction because it is the $x$-component of electronic velocity. This is the complexity and diversity of the information about particles that we have previously mentioned. In order for $\dot{x}$ to be meaningful, we have to sacrifice the electronic freedom of motion, namely, allow electrons to move only in $x$ direction in a straight line, because only in this way can $v_x$ be regarded as the speed of electrons. So the physical meaning of $\dot{x}$ can be explained as: $\dot{x}$ is the total velocity of a free electron that moves in $x$ direction in a straight line.

Since $v_{ox}$ and $v_x$ are respectively interpreted as the electronic oscillatory speed and speed, we can substitute $v_o$ for $v_{ox}$ and $v$ for $v_x$ in Eq.(8). And then we have

$$c^2 = v_o^2 + v^2 \tag{9}$$

As expected, the previously mentioned asymmetry has disappeared in Eq.(9).

It is clear that the total velocity of electrons equals to the velocity of light in Eq.(9). But we are not worried whether this result has violated special relativity because we will see that the oscillatory speed of a particle will never equal zero unless it is massless from Eq.(16), in other words, the electronic velocity is still less than the velocity of light. In reality, we can prove that Eq.(9) is fully compatible with the basic relations in special relativity.

Multiply both members of Eq.(9) by $(mc)^2$, we obtain

$$m^2 c^4 = m^2 v_o^2 c^2 + p^2 c^2 \tag{10}$$

Where $m$ is electronic mass and $p(=mv)$ is just the magnitude of electronic momentum. Apparently, the following equation has to be satisfied to make Eq.(10) the energy-momentum relationship, i.e.,

$$mv_o = m_0 c \tag{11}$$

Where $m_0$ is the rest mass of electrons. Substituting Eq.(9) in Eq.(11), we obtain without regard to the

negative energy solution

$$m = \frac{m_0}{\sqrt{1-(v/c)^2}} \tag{12}$$

Eq.(12) is just the well-known mass-velocity relationship. As expected, Eq.(10) is expressed as

$$m^2 c^4 = m_0^2 c^4 + p^2 c^2 \tag{13}$$

Eq.(13) is the energy-momentum relationship. Not only that, we find that the electronic rest energy all comes from the oscillation of the electron itself from Eq.(11), which gives the theoretical basis of which electrons are not composite particles, because the rest energy of a composite particle is commonly considered to be the sum of the kinetic energy and interaction potential of all the particles within it.

Now that Eq.(9) is fully compatible with the universally basic relations in special relativity, the particles respecting Eq.(9) are not limited to electrons. Therefore, we might as well take Eq.(9) as the definition of elementary particles. Its meaning can be explained as: The total velocity of a free elementary particle can only be the velocity of light. We do this according to the following considerations:

① An elementary particle certainly can not be a composite particle. As mentioned above, a particle can achieve that so long as it respects Eq.(9).

② An elementary particle must respect the basic relations in special relativity just as any other particles or bodies do. Nevertheless, only by respecting Eq.(9) can the particle with the total velocity respect Eq.(12) or Eq.(13).

③ In philosophy, dialectical materialism thinks that the physical world is in eternal motion, namely, the motion of matter is absolute. With the complexity and diversity of the existence of matter, the forms of motion are divided into several levels from simple to complex and the simplest one is mechanical motion. As the smallest units of matter, the motion of elementary particles is required to be both absolute and mechanical (the simplest). Such a motion can only be the motion related to light velocity, because the constancy of light velocity is only the absolute in mechanical motion.

Obviously, an electron is an example of elementary particles defined by Eq.(9), while a photon is another. The oscillatory speed of a photon equals zero due to its zero mass. According to Eq.(9), the total velocity of a photon can only be equal to its velocity — the velocity of light. So, a photon whose total velocity can only be the velocity of light, by definition, is an elementary particle, too.

In order to obtain the expansion of $v^2$ in Eq.(9), we rewrite Eq.(7) with the results of $\dot{x}$, $\dot{y}$ and $\dot{z}$, respectively.

$$c\alpha_j = v_{oj} + v_j \tag{14}$$

where

$$v_{oj} = \frac{i}{2}\hbar c \dot{\alpha}_j^0 e^{-i2Ht/\hbar} H^{-1} \tag{14a}$$

$$v_j = c^2 p_j H^{-1} \qquad j = x, y, z \tag{14b}$$

With the help of Eq.(14b), we obtain the expansion of $v^2$, i.e.,

$$v^2 = v_x^2 + v_y^2 + v_z^2 \tag{15}$$

where

$$v_j^2 = c^4 p_j^2 H^{-2} \qquad j = x, y, z \qquad (15a)$$

It is worth notice that $v_{oj}$ and $v_j$ appear symmetrically in Eq.(14) or Eq.(8). Could it be that the oscillation also has three components? That is true from the point of view of symmetry. As we all can see, Eq.(9) is obtained by the synchro replacements of $v_{ox}$ by $v_o$ and $v_x$ by $v$ in Eq.(8). Obviously, the symmetry of $v_{oj}$ and $v_j$ in Eq.(8) or similar equations would not be broken by such synchro replacements and thus can be well preserved in Eq.(9). In other words, $v_o^2$ ought to have the expansion similar to $v^2$.

The problem is that we are not yet able to write out the expanded form of $v_o^2$ directly, mainly because there is no additional condition when $v_{oj}$ is replaced by $v_o$, which means that all $v_{oj}$ would have the same eigenvalues. For instance, substituting the condition that electrons are allowed to move only in $j$ direction in a straight line in Eq.(6) ( relevant $x$ is substituted by $j$ ) and Eq.(14), we obtain

$$v_{oj} = m_0 c^3 \alpha_j \beta H^{-1} \qquad (16)$$

$$v_{oj}^2 = m_0^2 c^6 H^{-2} \qquad j = x, y, z \qquad (16a)$$

As expected, the eigenvalues of $v_{oj}$ have nothing to do with $j$ and all of them equal to the eigenvalue of $v_o$, because Eq.(16a) and Eq.(11) are actually the same for the particle which is in the eigenstate of energy. However, it reflects the fact that the components of oscillatory speed can not be observed. This gives us a chance to give $v_{oj}$ an eigenvalue by multiply $v_{oj}$ by a proper coefficient, i.e.,

$$V_{oj} = q_j m_0 c^3 \alpha_j \beta H^{-1} \qquad j = x, y, z \qquad (17)$$

where $q_j$ is a real number to ensure the hermiticity of the operator $V_{oj}$. And then we have

$$v_o^2 = V_{ox}^2 + V_{oy}^2 + V_{oz}^2 \qquad (18)$$

where

$$V_{oj}^2 = q_j^2 m_0^2 c^6 H^{-2} \qquad (18a)$$

and

$$q_x^2 + q_y^2 + q_z^2 = 1 \qquad (18b)$$

Frankly, we are almost ignorant of the expansion of $v_o^2$, and the only known is that $v_o$ is the superposition of $V_{oj}$. $v_o^2$ is nothing but a quadratic sum of $V_{oj}$ because the relations among all of other terms in Eq.(9)

has actually been quantitative rather than matrix's. The $q_j$ satisfies Eq.(18b) in order to enable Eq.(18) to be in conformity with Eq.(11).

It must be pointed out that these components of the oscillatory speed have nothing to do with the direction since we said the oscillation is isotropic. The real significance of which the oscillation has components is that an elementary particle can have several independent oscillations. Our concern at the moment is the differences between the particles that have a varying number of independent oscillations.

When a quantum system has more than one degree of freedom in quantum mechanics, its state function is often the product or direct product of wave functions corresponding to each degree of freedom. For example, we consider the state function of a free electron in the non-relativistic case, i.e.,

$$\psi(x,y,z) \otimes u_s = \exp(ip_x x/\hbar) \cdot \exp(ip_y y/\hbar) \cdot \exp(ip_z z/\hbar) \otimes u_s \qquad (19)$$

where $u_s$ is a spin wave function of two-component without regard to the negative energy solution, which corresponds to the intrinsic degree of freedom of the electron, and the symbol $\otimes$ represents the direct product. Obviously, $v_j$ operates on $\exp(ip_j j/\hbar)$, while $V_{oj}$ operates on $u_s$. This is because $V_{oj}$ is a real sense of matrices and has been reduced to a $2 \times 2$ matrix accordingly in the above cases. It is now clear that the oscillation is associated with the spin wave function. Since the spin wave function corresponds to one intrinsic degree of freedom, an electron has one oscillation. Then we might as well let $q_y = q_z = 0$ in Eq.(18b). Following this train of thought, it is easy to write out the state function of free elementary particles that have two or three independent oscillations, i.e.,

$$\exp(ip_x x/\hbar) \cdot \exp(ip_y y/\hbar) \cdot \exp(ip_z z/\hbar) \otimes u_{s1} \otimes u_{s2} \qquad (20)$$

or

$$\exp(ip_x x/\hbar) \cdot \exp(ip_y y/\hbar) \cdot \exp(ip_z z/\hbar) \otimes u_{s1} \otimes u_{s2} \otimes u_{s3} \qquad (21)$$

where $u_{s1}$, $u_{s2}$ and $u_{s3}$ are the spin wave functions of two-component and operators $V_{oj}$ operate on them one-to-one. Beside the direct product, the combination of $u_{s1}$, $u_{s2}$ and $u_{s3}$ can also be the form of the entangled state on the assumption that it is slightly more complicated. However, it will not affect the result of the following discussion.

In quantum mechanics, the direct product of two or more spin wave functions of two-component are generally used to describe the spin state of the system that is composed of corresponding quantity of spin-1/2 particles. Fortunately, we have known that the spin of quarks or antiquarks is 1/2 and they can only be found within hadrons. Therefore it stands to reason that quarks or antiquarks are regarded as the sources of these independent oscillations. And then the particles corresponding to Eq.(20) or Eq.(21) are associated with hadrons, specifically, the elementary particles defined by Eq.(9) in addition to photons are:
a) leptons corresponding to Eq.(19), whose spin is 1/2;
b) mesons corresponding to Eq.(20), whose spin is 0 or 1;
c) baryons corresponding to Eq.(21), whose spin is 1/2 or 3/2.

However, the particle that consists of more than three quarks, such as a tetraquark[2-5] or a pentaquark[6-9],

is ruled out because an elementary particle is allowed a maximum of three independent oscillations.

Like the fractional charges in the quark model, the quark here also have some peculiar properties:

① The oscillatory speed of a hadron comes from the superposition of that of the quarks within it in Eq.(18).
② Every quark has its own intrinsic degree of freedom but has no its own coordinate degrees of freedom in Eq.(20) or Eq.(21).

Obviously, a quark lacks the independence a particle should have. We therefore conclude that a quark is not a particle, but part of the hadron or a partial particle. Since both of the above properties can be the reason of quark confinement[10-12], quark confinement may serve as an evidence of this conclusion[13].

For instance, if a quark were isolated, it would be no longer a quark according to the property ②, because an isolated quark must have had its own independent coordinates. In addition, the oscillatory speed of an individual quark is only a part of that of the hadron according to the property ①, while the $v$ may serve as the speed of the quark because it is part of the hadron and has no its own independent coordinates. Since the total velocity of hadrons can only be the velocity of light, the total velocity of the quark is obviously less than the velocity of light, which means that an individual quark does not follow Eq.(9), as metioned earlier, and thereby not following the basic relations in special relativity. It is for that reason that quarks are kept within hadrons and prohibited from appearing in isolation.

---------------------------------